\newcommand{\simle}{\mbox{$\stackrel{<}{_{\sim}}$}}

\newcommand{\secd}{\,{\rm s}}
\newcommand{\km}{\,{\rm km}}

\newcommand{\kms}{\km\secd^{-1}}

\documentstyle[11pt,aaspp4,flushrt]{article}
\lefthead{Monnier et al.}
\righthead{Spectra of late-type Stars}

\begin{document}
\title{Mid-infrared spectra of late-type stars: Long-term evolution}
\author{J. D. Monnier\altaffilmark{1}, T. R. Geballe\altaffilmark{2}, 
and W. C. Danchi\altaffilmark{1}
} 
\altaffiltext{1}{Space Sciences Laboratory, University of California, Berkeley,
Berkeley,  CA  94720-7450
}
\altaffiltext{2}{Gemini Observatory,
670 North A'ohoku Place, University Park, Hilo, HI 96720}

%
%
%
\begin{abstract}
Recent ground-based mid-infrared spectra of 29 late-type stars, most
with substantial dust shells, are compared to ground-based spectra of
these stars from the 1960s and 1970s and to IRAS-LRS spectra obtained
in 1983.  The spectra of about half the stars show no detectable
changes, implying that their distributions of circumstellar material
and associated dust grain properties have changed little over this
time interval.  However, many of the stars with strong silicate
features showed marked changes.  In nearly all cases the silicate peak
has strengthened with respect to the underlying continuum, although
there is one case (VY~CMa) in which the silicate feature has almost
completely disappeared.  This suggests that, in general, an
oxygen-rich star experiences long periods of gradual silicate feature
strengthening, punctuated by relatively rare periods when the feature
weakens.  We discuss various mechanisms for producing the changes,
favoring the slow evolution of the intrinsic dust properties (i.e.,
the chemical composition or grain structure).  Although most IRAS
spectra agree well with ground-based spectra, there are a number of
cases where they fall well outside the expected range of uncertainty.
In almost all such cases the slopes of the red and blue LRS spectra do
not match in their region of overlap.
\end{abstract}


\section{Introduction}
Although the first mid-infrared (5-20\,$\micron$) spectra of late-type
stars were taken over thirty years ago, no systematic comparison with
more recent spectra of identical sources has ever been made.  In the
late 1960's and early 1970's, Merrill, Stein, and collaborators
obtained a homogeneous set of mid-infrared spectra on a large set of
oxygen-rich and carbon-rich asymptotic giant branch (AGB) stars.  In
the 1980's, the Infrared Astronomical Satellite (IRAS) provided an
atlas of 8-22\,$\micron$ spectra, but on the question of comparison
with previously observed sources the IRAS Explanatory Supplement
merely stated: ``The comparison showed satisfactory results
(\cite{iras88}, pp. IX-20).''  Recently, we published mid-infrared
spectra of red giants and supergiants, many of which show long-period
variability, with increased temporal coverage to resolve spectral
changes taking place on a pulsational time scale (\cite{mgd98}=MGD98).

MGD98 identified stars that showed significant spectral shape
variations during a pulsational cycle and characterized the observed
changes.  Some observers in the past have interpreted spectral changes as
reflecting long-term changes rather than pulsational ones (e.g.,
\cite{fgs75}), but their data did not adequately sample a sufficiently
long time interval to test this interpretation.  Cognizant of these
issues, we now compare these three homogeneous sets of data to search
for significant long-term changes in spectral shape.

The 8-13\,$\micron$ spectra of most of these stars contain significant
emission from circumstellar dust grains.  If this dust is forming in
steady state at a typical condensation temperature of $\sim$1200\,K,
then the inner dust radius is at $\sim$3-4~R$_\star$, corresponding to
about 10\,AU for Mira variables.  Most of the mid-infrared dust
emission comes from this region containing the hottest dust.  If the
dust is moving at typical velocities for molecules observed in the
outflows ($\sim 10 \kms$) then the emitting dust within $\sim$20~AU is
replenished every 10~years.  Hence, we expect an investigation of
mid-infrared spectra over a 25~year time span to effectively probe the
temporal uniformity of the mass-loss around a given star.

In this paper we first introduce and describe the data sets used for
spectral comparison.  Care was taken to understand the particular
calibration procedures and most of the spectra were extracted from
large homogeneous sets.  Also by comparing these sets of data
as a whole, systematic biases in any given data set may be discovered
through comparison with the other two.  Indeed, in our comparison of the data
sets we have discovered a subset of miscalibrated LRS spectra in the IRAS
database.  Lastly, we discuss the implications of our results for the
understanding of mass-loss processes on the AGB and single out a few stars
for special comment.

\section{The Data Sets}
This paper makes use entirely of published mid-infrared spectra, and
we refer the interested reader to the original articles for complete
observing details.  For most stars considered here, three epochs of
data exist in the literature: one from the late 1960's or early 1970's
(Epoch I), one from the IRAS-LRS Atlas observed in 1983 (Epoch II),
and, lastly, UKIRT data from the mid-1990's (Epoch III).  See Table~1
for additional information on the target stars.  We will now describe
the data set used for each epoch and the calibration issues unique to
each.

Merrill \& Stein (1976abc) published the most extensive set of
mid-infrared spectra during the late 1960's and 1970's and we have
used their data when available.  This strategy makes the comparison of
spectra taken decades apart less susceptible to varying calibration
techniques of the observers.  Since all spectra by a given observer
may suffer from the same systematic errors, these errors may be
identified by comparison with a more recent, homogeneous set of
spectra.  In addition, spectra from Gillett, Low \& Stein (1968),
Hackwell (1972), Forrest, Gillett \& Stein (1975), and Giguere, Woolf,
\& Webber (1976) were used to supplement the 3-epoch data set
available for analysis.  All these authors calibrated their spectra by
observing standard stars and assuming blackbody temperatures (e.g.,
$\alpha$~CMa as a 10000\,K blackbody), although the calibration star
used for each particular spectrum was rarely documented.
Occasionally $\alpha$~Boo (K1.5III) and $\alpha$~Tau (K5III) were
used, being approximated as 4000\,K blackbodies
(\cite{gf73}), a problematic assumption due to absorption in the
fundamental band of SiO (\cite{cohen95}).  However, recent work
(\cite{cww92}) has found that the spectrum of $\alpha$~Tau
can be fit very well by a 3800\,K blackbody between 8 and
20\,$\micron$, and hence these calibrations made by Merrill, Stein,
and others are quite satisfactory in this wavelength regime.
 
The InfraRed Astronomical Satellite (IRAS) was launched in January
1983 and operated for nearly a year.  It was outfitted with the
Low-Resolution Spectrometer (LRS) which collected thousands of spectra
at wavelengths between 8 and 22\,$\micron$.  The spectrometer measured
two overlapping wavelength ranges: one from 7.7 to 13.4\,$\micron$ and
another from 11.0 to 22.6\,$\micron$.  The spectral shape was
calibrated by assuming the intrinsic $\alpha$~Tau spectrum was equal
to a 10000\,K blackbody; this and additional information can be found
in the IRAS Explanatory Supplement (1988).  Cohen, Walker \& Witteborn
(1992) measured the mid-infrared spectrum of $\alpha$~Tau
independently and found systematic deviations from the ideal Planck
spectrum, and their correction factors were applied to all IRAS-LRS
data before comparison in this paper.  The IRAS-LRS Atlas data were
obtained from an internet repository maintained by Kevin Volk at the
University of Calgary
(http://iras3.iras.ucalgary.ca/\verb+~+volk/getlrs\_plot.html) and
further description of the data set, including scans not originally
included in the IRAS-LRS Atlas, can be found elsewhere (\cite{vc89};
\cite{volk91}; \cite{kvb97}). 

The most recent epoch of mid-infrared spectrophotometry considered
here was carried out at the United Kingdom Infra-Red Telescope
(UKIRT) from 1994 to 1997 (\cite{mgd98}).  All standard stars used for
flux calibration were of spectral type K0 or earlier, so that
absorption in the fundamental band of SiO, which is very prominent in
late K and M giants and supergiants, is minimal in the ratioed spectra
and does not affect the shapes of the reduced spectra.

In this paper we do not concern ourselves with the absolute flux level of 
the mid-infrared spectra, because of the difficulty in reconciling the
broad-band filter bandpasses used by IRAS and the early workers, which
would be required to precisely compare the absolute
photometry.  In addition, previous workers sampled the stellar spectrum at
incongruous pulsational phases.  For example, Merrill \& Stein (1976ab)
presented relative spectrophotometry accompanied by a table of broad-band
fluxes; unfortunately, spectra obtained at various and undocumented phases
during the pulsational cycle of variable stars were averaged together.
This is also true for the IRAS-LRS spectra; the published Atlas
spectra result from an averaging of multiple scans taken at different times.

Therefore, we present only relative spectra, normalized to the average
broadband 8-13\,$\micron$ flux.  Only the recent UKIRT data
(\cite{mgd98}) sampled the stellar spectra throughout the pulsational
cycle, and we include spectra nearest to maximum and minimum for all stars
when possible.  Although adding some clutter to the figures, this
allows one to determine if spectral shape differences are likely due
to pulsation effects, rather than long-term evolution of the dust
shell properties.

Observers during Epochs~I and III generally used photometric apertures
about 5\arcsec\, in size, while IRAS had a larger one
(6\arcmin$\times$15\arcmin).  The Egg Nebula is the only source in
this paper whose mid-infrared emission is known to be extended on
the former scale.  The Infrared Spatial Interferometer has surveyed a
number of these bright mid-infrared sources (e.g., \cite{danchi94})
and showed that nearly all of the mid-infrared emission occurs on
scales smaller than 5\arcsec.  However, direct imaging of faint
emission around these stars has rarely been done.  In a few cases,
some emission has been detected outside of the small aperture used by
Epoch I and III observers.  In particular, Sloan and collaborators
have detected distant silicate material around $\alpha$~Ori
(\cite{sloan93}) and carbon grains around IRC~+10216 (\cite{sloan95}).
While only contributing a small fraction of the total flux, it is
possible that such missing flux can contribute significantly to
spectral features in the mid-infrared region in some cases.  Hence some
small differences observed between Epoch II (IRAS) and the other data sets
may be attributable to such an effect.

\section{Results}

\subsection{Categorization Procedure}
In order to avoid the ambiguity associated with subjective
classifications, a quantitative measure of spectral variability was
developed to identify sources exhibiting statistically significant
long-term evolution of their mid-infrared spectral shapes.  A simple
$\chi^2$-based statistic, defined in the Appendix, allows robust
identification of stars exhibiting long-term changes in their spectra.
This statistic takes into account the pulsation-related spectral changes observed in MGD98 as well as measurement error in determining the likelihood of
long-term spectral variability.

Using the procedure outlined in the Appendix, two relevant statistics,
$\chi^2_{\rm{Epoch\,I}}$ and $\chi^2_{\rm{IRAS}}$ have been calculated
for all sources, and these results appear in Table\,1.  These values
are indicative of the spectral changes occurring between Epoch I and
Epoch III and between Epoch II (IRAS) and Epoch III, respectively.
Epoch I and Epoch II data were not quantitatively compared
to each other because the spectra did not temporally sample the
pulsational cycle of the sources.

The stars in this survey showed a continuum of variability; some stars
evinced virtually no variability, while others showed obvious spectral
changes.  In order to divide our sample, we have adopted the cutoff
value of $\chi^2 = 1.5$ to indicate significant variability.  Based on
these quantities, the stars have been divided into three categories
and presented in Figures~1-3.  Figure~1 contains the spectra of
6~stars which are consistent with the hypothesis of no spectral
evolution between all three epochs.  Figure~2 contains spectra of
9~stars that show good agreement between Epoch~I and Epoch~III, but
in which the IRAS spectra appear discrepant.  Finally, Figure~3
presents 14~stars whose spectral shapes are significantly
different between Epoch I and Epoch III, independent of the IRAS data.

\subsection{Constant Stars}
Figure~1 contains spectra of stars whose spectral shapes have not
changed, within known uncertainties, between all three epochs under
study.  Quantitatively, the selection criterion for inclusion in this
figure is $\chi^2_{\rm Epoch~I}\leq 1.50$ and $\chi^2_{\rm IRAS}\leq
1.50$; in addition, three stars with no available Epoch I data are
included here.  Many of these stars are calibrators because of
their general lack of variability and relatively little dust emission
(e.g., $\alpha$~Boo, $\alpha$~Tau, $\delta$~Oph).  The excellent
agreement found for these stars provides confidence that changes in
spectral shape described in the next two sections are indeed real and
not the result of the differing calibration procedures of the
individual observers.  Note that CIT~3 is highly variable, and
the large spectral shape changes occurring during its pulsational cycle
would mask all but the most dramatic long-term spectral evolution.

\begin{figure}
\begin{center}
\centerline{\epsfxsize=\columnwidth{\epsfbox{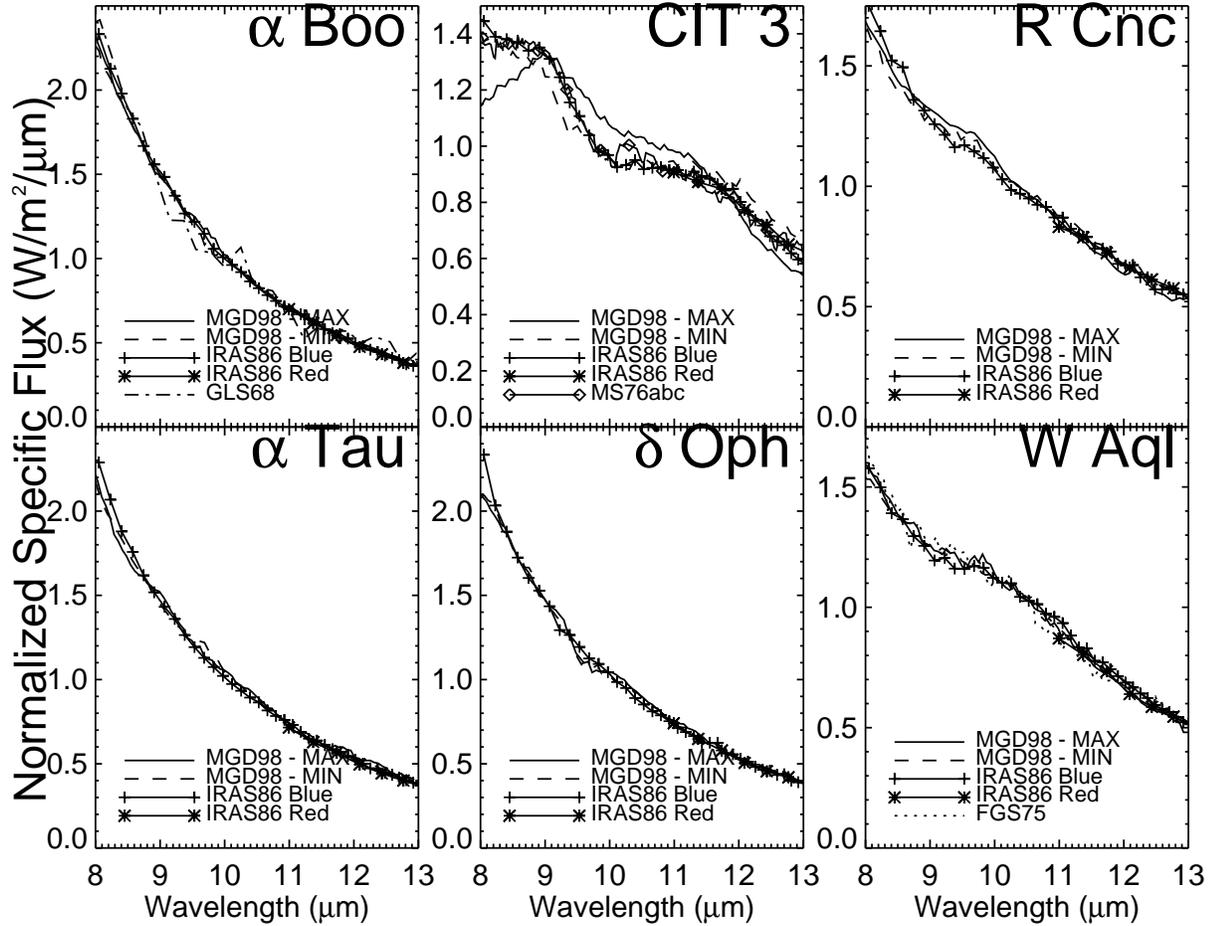}}}
\caption{Normalized mid-infrared spectra of late-type stars that showed no
apparent changes in spectral shape using data from Epochs I, II, and III.
Epoch III spectra near maximum and minimum flux are shown.
Spectra references are for
Gillett, Low, \& Stein (1968=GLS68), Hackwell (1972=Hackwell72),
Forrest, Gillett, \& Stein (1975=FGS75), 
Merrill \& Stein (1976=MS76abc), IRAS Science Team (1986=IRAS86),
and Monnier, Geballe, \& Danchi (1998=MGD98).
}

\end{center}
\end{figure}

\subsection{IRAS data}
Figure~2 presents spectra of stars that show excellent agreement
between the earliest and the most recent epochs ($\chi^2_{\rm Epoch~I}
\leq 1.5$), but whose IRAS measurements show systematic disagreement
($\chi^2_{\rm IRAS} > 1.5$) with recent data.  In a few obvious
cases, the miscalibration of one of the IRAS-LRS detectors appears
responsible (e.g., $\alpha$~Her) for this discrepancy.  The red and
blue detectors of IRAS-LRS share a common wavelength region between
11.1 and 13.4\,\micron, and should report self-consistent spectral
fluxes.  In the Appendix, we define the statistic $\sigma_{\rm
overlap}$ which quantifies how well the observed spectral slopes in
the two LRS detectors agree in the overlap region.  In short, a
straight line is fitted to the ratio of the blue and red data in the
overlap region.  For self-consistent data, the slope should be zero
and $\sigma_{\rm overlap}$ is the number of standard deviations away
from zero of the best-fit slope.
Hence, values above 3 are very
unlikely ($<1$\%) to occur by chance for well-calibrated data and
Gaussian random noise.  This statistic has been calculated for all
IRAS data in this paper and the results can be found in Table~1.

\begin{figure}
\begin{center}
\figurenum{2}
\centerline{\epsfxsize=\columnwidth{\epsfbox{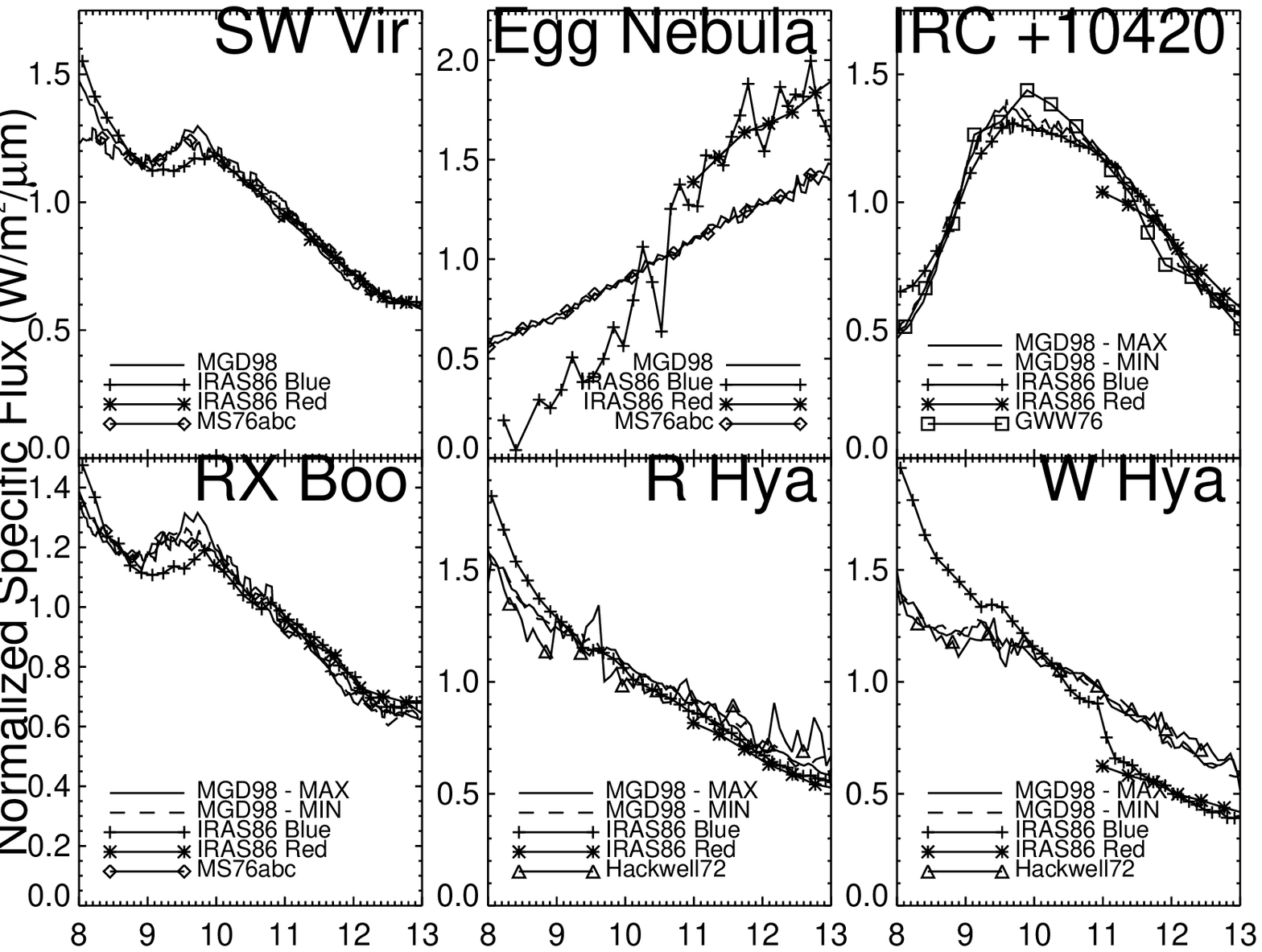}}}
\caption{Normalized mid-infrared spectra of late-type stars that showed no
apparent changes in spectral shape between Epochs I and III, but whose
IRAS spectra (Epoch II) appear discrepant.
Refer to Figure~1 for spectral references.
}

\end{center}
\end{figure}

\begin{figure}
\figurenum{2}
\begin{center}
\centerline{\epsfxsize=\columnwidth{\epsfbox{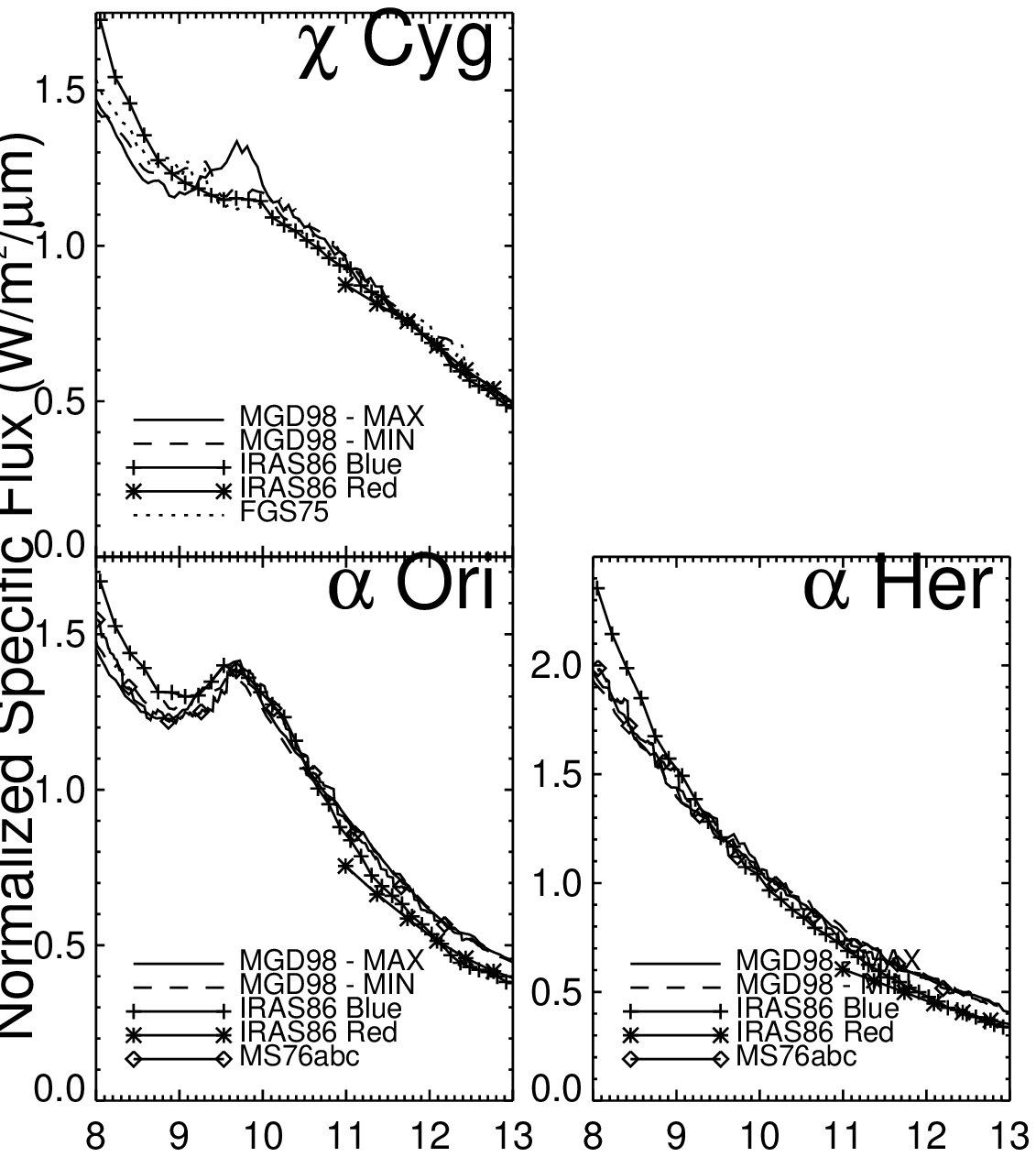}}}
\caption{continued}
\end{center}
\end{figure}

The spectra in this figure are ordered by increasing $\sigma_{\rm
overlap}$; hence, spectra at the end of the figure show this
miscalibration effect most strongly.  There is no evidence for
miscalibration problems in the overlap region ($\sigma_{\rm
overlap}\simle2.0$) for the first three spectra (SW~Vir, RX~Boo, Egg
Nebula).  The large angular size of the Egg Nebula is the most likely
explanation for the different spectral shapes observed for that star
(see previous discussion at the end of \S2), However, RX~Boo and
SW~Vir are not known to be extended, and the apparent spectral
variability may indicate the presence of transient
strengthening/weakening of the silicate feature around these two
Semi-regular variable stars.  Note that only one Epoch III spectrum exists
for SW~Vir, hence the importance of pulsation-related
variability on this star's spectral shape is not known.  In fact, the formal
$\chi^2_{\rm Epoch~I}$ for SW~Vir is larger than 1.5, but was included
in this figure because differences between the Epoch I and Epoch III
data are likely due to pulsation effects not quantified by the
statistic.

By ordering the spectra by $\sigma_{\rm overlap}$, we can see a
systematic spectral ``tilt'' associated with the spectral
slope-mismatch.  All the IRAS spectra in Figure~2 with large
$\sigma_{\rm overlap}$ also evince an overall spectral shape which is
{\em bluer} than the Epoch I and III data, easily seen as excess
emission between 8 and 9\,\micron.  Hence, it is reasonable to conclude
that most of the IRAS-LRS data with high $\sigma_{\rm overlap}$ values
have some residual miscalibration, not previously identified, and
appear slightly bluer than in reality.  Inspection of Table~1 reveals
that 11 of the 29 IRAS-LRS spectra considered here have $\sigma_{\rm
overlap}\geq 3.0$.  An understanding and future correction of
this miscalibration may be useful.

\subsection{Evidence for long-term changes}
The spectra of the remaining 14~stars appear in Figure~3a-c, and all show
significant changes between Epochs~I and III.  The Epoch~II spectra of
some of these stars are flawed in the same manner as discussed in
\S3.3, while others appear to be largely consistent with one epoch or
the other, or intermediate between the two.  All spectra here have
$\chi^2_{\rm Epoch~I}>1.5$ and the $\chi^2_{\rm IRAS}$ statistic was
not used in determining membership in this class.
In addition, these 14 stars have been further subdivided based on the type of
spectral variability observed.  

\subsubsection{Strong silicate emission features}
Figure 3a contains the spectra of 7 stars whose strong silicate features
show long term variability.
Because the dust grains around these
stars have such strongly wavelength-dependent 
opacities near 10~$\micron$, the profile of the silicate
feature and its equivalent width are highly sensitive to changes in the
spectral emission characteristics, temperature, and optical depth structure
of the dust shells.  Many of these stars were identified in MGD98 as
possessing silicate features that change shape even during a pulsational
cycle.  However, the observed changes between Epoch~I and III lie outside the
envelope of variation previously observed.

\begin{figure}
\begin{center}
\figurenum{3}
\centerline{\epsfxsize=\columnwidth{\epsfbox{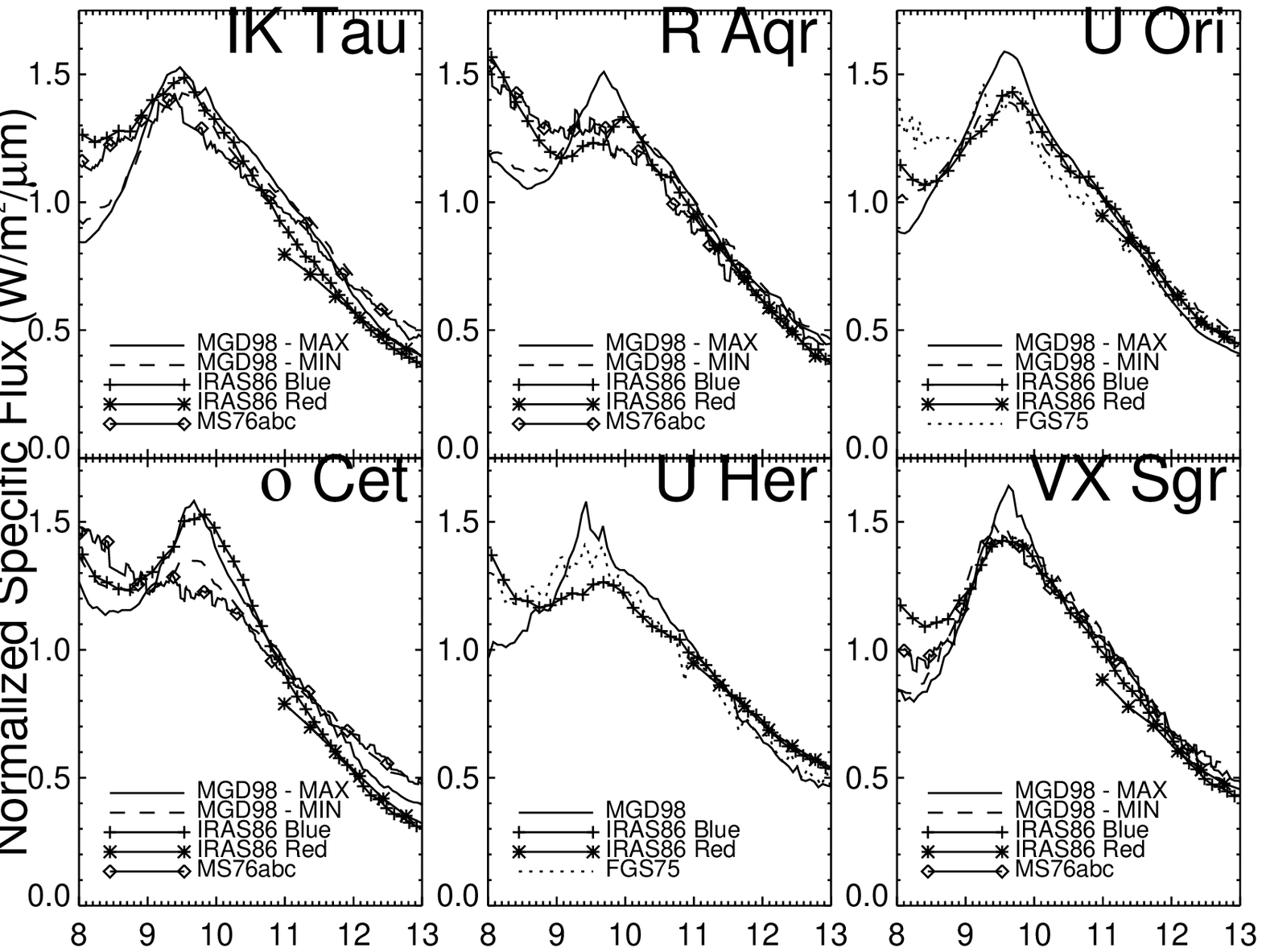}}}
\caption{Normalized mid-infrared spectra of late-type stars that showed 
significant changes in spectral shape between Epochs I and III, independent
of the IRAS data. Refer to Figure~1 for spectral references.
{\em a.} Stars with strong silicate emission features. }
\end{center}
\end{figure}

\begin{figure}[t]
\begin{center}
\figurenum{3a}
\centerline{\epsfxsize=\columnwidth{\epsfbox{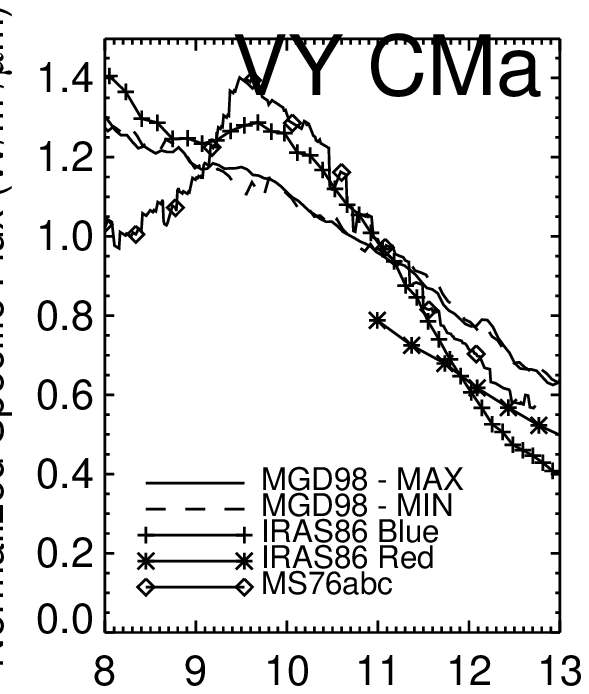}}}
\caption{
continued}
\end{center}
\end{figure}

The changes observed in the emission spectra of the dust could be due to
long-term variations in the properties of the dust particles
themselves.  Dust optical constants can be affected by changing
chemical abundances or other physical conditions in the stellar
atmospheres (see MGD98) as the star evolves, perhaps linked to thermal
pulses or other nuclear shell burning phenomena.  Changes in the
elemental composition may be the only way to explain the dramatic
change in the VY~CMa spectrum (see below).

Alternatively, these differences may be due to episodic mass-loss
events, the observed variations in the 
spectra being consistent with an overall cooling
of the circumstellar dust shell.  However, interferometric
observations of these dust shells in the infrared indicate the
presence of dust close to the expected condensation radius
(\cite{dyck84}; \cite{danchi94}), making it difficult to understand
why the mean dust shell temperatures should all be decreasing in time.
See MGD98 for more detailed discussions on the dust temperature's
effect on the mid-infrared spectra.  In short, the observed changes in
the spectral
shape require a large temperature change which would imply a
dust-star separation incompatible with interferometer measurements.

Regardless of the mechanism for producing the observed changes, the
fact that few stars showed a decrease in the silicate feature peak
relative to the continuum suggests that these stars spend most of
their time building up their silicate peaks.  Assuming the silicate
strengthening has been gradual and continuous over the last 25 years,
the typical time between episodes of silicate feature weakening must
be much longer than that probed here.  Considering
that 5 Mira variables with the strongest silicate emission 
features all showed the same
systematic change in spectral shape, we estimate that this time
scale is greater than about 125~years.  Because of the fewer number of
supergiant candidates, we can only say the time scale relevant for
these stars is likely to be greater than $\sim$40~years.

\subsubsection{Spectral slope changes}
Figure 3b shows the spectra of 4 stars whose Epoch I and Epoch III
spectra differ in slope by an amount well outside that expected from the
statistical error in the measurements.  
In all cases here, the spectra have become redder, consistent with a
long-term cooling of the dust shell.  However, the slopes of the Epoch~II
spectra are not always intermediate
to the Epoch I and III
spectra, and hence the spectral tilt may have different origin (e.g.,
R~Leo).  Variation in the dust production on time scale of 10-20 years
could produce such changes, and would be in general agreement with
conclusions based on interferometry (e.g.,
\cite{danchi94}).  However, we note that a slight spectral
miscalibration across the wavelength band can also mimic this
effect.

\begin{figure}
\figurenum{3}
\begin{center}
\centerline{\epsfxsize=\columnwidth{\epsfbox{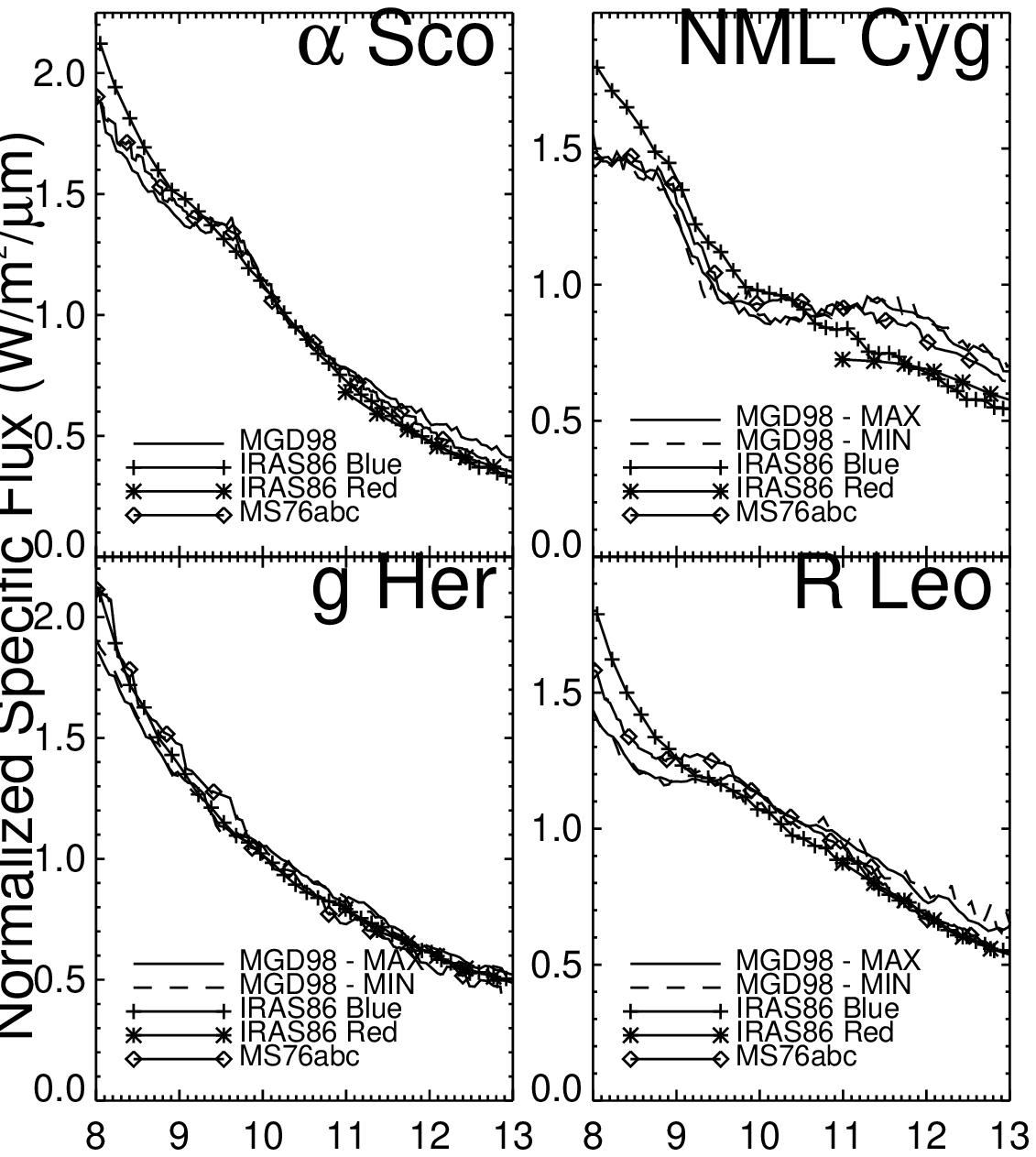}}}
\caption{{\em b.} Stars evincing long-term spectral tilts.
}
\end{center}
\end{figure}

\subsubsection{Excess variability of the carbon stars}
The spectra of the 3 carbon stars in our sample are found in Figure 3c.
MGD98 noted that the spectra from the carbon stars showed more 
apparent variability on a pulsation time scale than oxygen-rich
stars with weak silicate features.  The long-term variability reported here
is also fairly small, but statistically significant.  Variability of the complex dust distributions recently observed around a few such
stars using interferometric techniques 
(\cite{hb98}; \cite{weigelt98}; \cite{tmd99}) may play a role in this observed
enhanced variability.

\begin{figure}[h]
\figurenum{3}
\begin{center}
\centerline{\epsfxsize=\columnwidth{\epsfbox{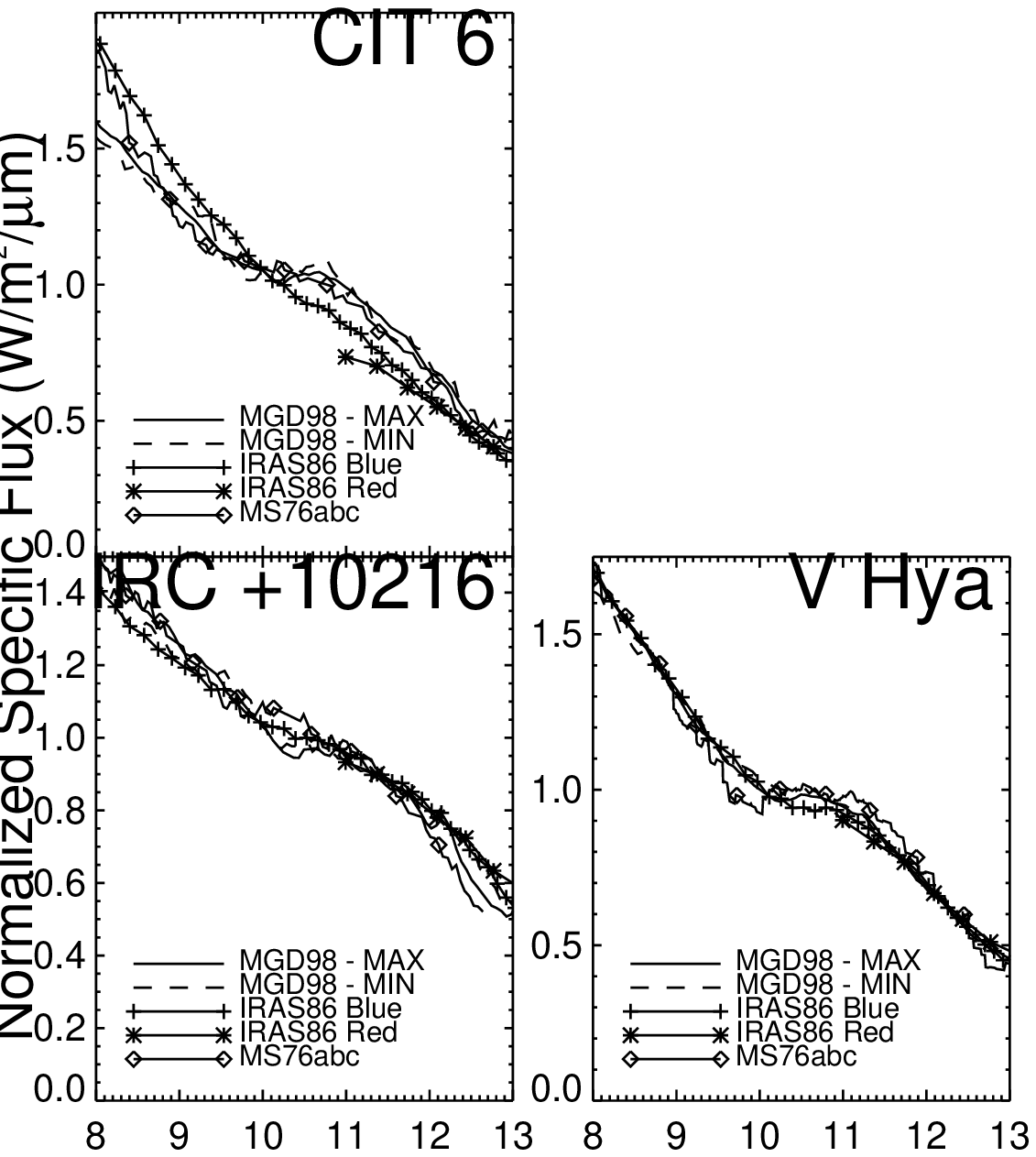}}}
\caption{
{\em c.} Carbon stars, displaying enhanced variability.
}
\end{center}
\end{figure}

\subsection{Comments on individual sources}

{\em $\alpha$~Ori:} 
The silicate feature has shifted toward slightly shorter wavelengths
now.  While not consistent with a generally cooling of the dust shell,
this may be caused by slight replenishing of inner dust shell as
recently observed by Bester et al. (1996).  Although poor calibration could
explain the difference, changes in the optical properties of the dust
as it ages may be important here.

{\em Egg Nebula:} 
Observers during Epochs~I and III generally used
photometric apertures about 5\arcsec\, in size, while IRAS had a
larger one (6\arcmin$\times$15\arcmin).  Because this nebula is
extended on the same scale as the smaller, ground-based apertures,
direct comparison of the data sets is difficult. 

{\em IRC~+10420:}
The Epoch I data for this source is rather noisy and was included only to
demonstrate that no gross changes in the silicate feature have 
occurred (see VY~CMa comments below).   The high noise level prevents
a detailed comparison between Epoch I and II data sets.

{\em R~Hya:}
Peculiarly, the IRAS-LRS Atlas spectra from the blue and red detectors
agree well in slope but are noticable offset from one another (see Figure~2).

{\em U Her:} 
Because MGD98 only published a single observation of this
source, the identification of this source as experiencing a long-term
spectral variation is suspect.  However, the Epoch I observation
appears visually to be outside the typically envelope of
pulsation-related variability seen in the other spectra of Figure~3a.
New observations of this source during a pulsational cycle are
important to resolve this ambiguity.

{\em VY~CMa:} 
The mid-infrared spectrum of the red supergiant VY~CMa has undergone a
remarkable change over the last 25 years (\cite{mgd98}).  What was once one
of the strongest silicate features observed by Merrill \& Stein (1976a) is
now nearly flat and featureless.  Although the spectral slopes of the
IRAS-LRS red and blue detectors are significantly different, it appears that
the strength of the feature was intermediate in 1983 to that observed before
and after and suggests the spectral change did not occur ``overnight,'' but
instead evolved over a number of years.  A significant increase in 
optical depth, sufficient to cause self-absorption in the silicate 
feature, could explain most of the flattening of the spectrum.  
However, a self-absorbed silicate feature usually shows more residual structure
(e.g., NML~Cyg, CIT~3) than apparent here, implying the emission
properties of the dust grains themselves have undergone a relatively rapid
transformation.
It is not known what physical mechanism
could cause such a dramatic change in the dust properties, but perhaps could
be due to a change in the chemical abundances.  This may be related to
the ``remarkable change'' in the infrared polarization between 1970 and 1974
observed by Maihara et al. (1976).  See Monnier et al. (1999) for recent
high-resolution observations which sheds some light on this mystery.

\section{Conclusions}
We have compiled a multi-epoch set of mid-infrared spectra on 29~stars
spanning a $\sim$25~year observing period.  This comparison has
established the high quality of published ground-based spectra and
allowed an investigation into the long-term stability of dust shells
and their properties.

The mid-infrared spectral shapes of about half of the target
stars have remained nearly constant over the entire period of time.
This suggests that the mass-loss rates have not changed significantly
during the last 25~years.  This constancy has allowed us to show  
that IRAS-LRS spectra may be somewhat miscalibrated when the slopes of the
red and blue spectra do not match in their shared wavelength region.

Most of the stars showing strong silicate features have experienced an
increase in the strength of the silicate peak relative to the
continuum.  This is naturally explained if an oxygen-rich star generally
experiences long periods of gradual silicate feature strengthening,
punctuated by relatively rare periods when the feature weakens.  The
physical conditions regulating dust formation may be varying on this
long time scale, affecting the optical properties of the dust
particles in the outflow.  Alternatively, general cooling of the dust
shells over the last 25~years can explain this effect, although this
hypothesis is at odds with interferometric observations of hot dust
around many of these stars.  The dramatic near-disappearance of the silicate 
emission feature around VY~CMa deserves attention, and may lead to new insights
into mass-loss processes around evolved stars.

This rudimentary analysis highlights the not surprising fact that the
circumstellar environments of late-type stars are not always simple, and
that assumptions of steady-state dust production on time
scales of decades can not always be justified.  These data provide further
evidence that mass-loss characteristics and dust signatures change not only
on the time scale of the large amplitude pulsations, but on
significantly longer ones as well.

\acknowledgments
{The authors would like to thank Kevin Volk for providing IRAS-LRS
spectra via his internet repository and for valuable discussion of its
contents.  JDM thanks C. D. Matzner and A. P. Zemgulys for 
numerous discussions.  
This research has made productive use of NASA's Astrophysics Data 
System Abstract Service.
This analysis was supported by the National
Science Foundation (Grants AST-9315485 \& AST-9731625)
and by the Office of Naval Research (OCNR N00014-89-J-1583 \&
FDN0014-96-1-0737).

}
\appendix
\section{Definition of $\chi^2$ and $\sigma_{\rm{overlap}}$ statistics}

The analysis in the body of this paper has made use of a few simple
statistics to evaluate the constancy of stellar spectral shapes
between epochs widely separated in time.  The purpose of this Appendix
is to precisely define the parameters $\chi^2$ and
$\sigma_{\rm{overlap}}$ used in the text.  Since only the spectral
{\em shapes} are under scrutiny here, all comparisons are done between spectra
which have been normalized by their mean fluxes.  

\subsection{Definition of $\chi^2$}
The $\chi^2$ statistic is simply a measure of how well a previous
spectra {\em fits} a recent one.  The $\chi^2$ defined here deviates
from the classical one in two respects.  Firstly, we 
report the reduced $\chi^2$, i.e. the $\Sigma\chi_i^2$ per degree of freedom,
instead of the $\Sigma\chi_i^2$;
hence, values greater than unity indicate a poor fit.  Secondly, the
uncertainty in the recent spectra (MGD98) is a combination of
pulsation-related variability and measurement error.  A uniform 2.0\%
relative uncertainty was assumed for all MGD98 data points (as estimated in
the original paper).  However, if the spectral shapes of MGD98 data at
flux maximum and minimum showed greater variation than this due to
pulsation-related effects, then the observed variation was used to
indicate the spectral shape uncertainty.

If we denote the normalized
MGD98 spectra at flux maximum and minimum by $S^{\rm{max}}_\lambda$ and 
$S^{\rm{min}}_\lambda$, then the wavelength-dependent uncertainty, $\delta S_\lambda$,
is defined below, where $S^{\rm{mean}}_\lambda=\frac{1}{2} [S^{\rm{max}}_\lambda 
+ S^{\rm{min}}_\lambda ]$,

\begin{equation}
\delta S_\lambda = \rm{Maximum}\left(
 \frac{1}{2} [S^{\rm{max}}_\lambda 
- S^{\rm{min}}_\lambda ] , 
0.02 \times S^{\rm{mean}}_\lambda\right).
\end{equation}

Measurement error for the previous epoch data, $\sigma_\lambda$, was
estimated from the scatter of data values over small wavelength
intervals. These sources of uncertainty were combined in quadrature,
and the reduced-$\chi^2$ evaluated as prescribed below. The averaging
was done over the wavelength range 8.1 to 12.5\,$\micron$, slightly
smaller than the fiducial bandpass due to limited coverage of a few
Epoch I spectra.

\begin{equation}
\chi^2_{\rm{Epoch\,I}}={\left<   \frac{ [S^{\rm{mean}}_\lambda  - 
S^{\rm{Epoch\,I}}_\lambda]^2} { (\delta S_\lambda)^2 + 
(\sigma^{\rm{Epoch\,I}}_\lambda)^2 } \right>}_\lambda .
\end{equation}

Although the above equation was written for Epoch~I, the $\chi^2_{\rm{IRAS}}$ 
statistic has the identical form.

\subsection{Definition of $\sigma_{\rm{overlap}}$}
This statistic was used to identify IRAS-LRS Atlas spectra whose blue
and red detector results were inconsistent in their common,
overlapping wavelength region, between 11.0 and 13.1\,$\micron$.  As
with the previously defined $\chi^2$ statistic, we chose a very simple
statistical criterion.

By restricting ourselves to the overlap region defined above, we can
define a wavelength-dependent ratio by dividing the data from the blue
detector by the data from the red detector.  Self-consistency would
demand that this ratio be equal to unity for all values in the
wavelength overlap region (within the noise).  However, it was
apparent from the data (e.g., $\alpha$~Her) that occasionally the blue
detector displayed a steeper slope than that observed by the red
detector.  This effect caused the ratio of the blue to red values to
have a slope across the wavelength overlap region.  A line was fit
(using the least-squares criterion) through the ratio vs. wavelength
curve data, and parameter uncertainties were estimated using
standard bootstrap-sampling techniques.  This produced an estimate of
the line slope and its standard deviation.  The
$\sigma_{\rm{overlap}}$ parameter is simply the magnitude of this
slope divided by the standard deviation.  For perfectly calibrated
data and Gaussian noise, more than 99\% of the slope determinations
should be within 3 standard deviations of 0.0.  An identical analysis
performed on simulated data indeed confirmed that we statistically
expect fewer than 1 of 29 fits to be outside of 3 standard deviations.
However, 11 of the 29 spectra (38\%) surveyed here were found with
$\sigma_{\rm{overlap}}$ greater than 3.0 (see Table~1), indicating
that spectral slope mismatches in the overlap region do affect a
significant fraction of the IRAS data considered.


 \pagebreak


\begin{deluxetable}{cllcclllc}  
\small
\tablewidth{0pt}
\tablecaption{Stellar Characteristics}
\tablenum{1}
\tablehead{
\colhead{Figure} & \colhead{Name} & \colhead{Alternate} & \colhead{Spectral} & \colhead{Variable} &  \colhead{$ {\chi}^2_{\rm{Epoch\,I}}$} &
\colhead{$ {\chi}^2_{\rm{IRAS}}$} & \colhead{$ {\sigma}_{\rm{overlap}}$} &
\colhead{Epoch I} \\
\colhead{Number} &		& \colhead{Names}     & \colhead{Type\tablenotemark{a}} & \colhead{Type$\tablenotemark{a}$} &
& & & \colhead{Reference}  			 
}

\startdata
1 &$\alpha$ Boo	& & K1.5III  	     & 			  &1.00&0.37&0.01& 1 			 \\
&$\alpha$ Tau	& & K5III		     &		 	&n/a&0.83&1.43& n/a 			 \\
&CIT 3 & WX~Psc& M9 &  Mira		         &0.84&1.50&3.35& 2 			 \\
& &IRC+10011 &&&&&&\\
&$\delta$ Oph    & &M0.5III            &                    &n/a&0.74&1.12&n/a                   \\
&R Cnc		& &M7IIIe	     &  Mira		  &n/a&0.91&1.00& n/a	 \\
&W Aql		& &S4.9		     &  Mira		  &0.61&0.42&2.14& 4 \\
\\
2&SW Vir & IRC+00230 &M7III		     &  SRb		  &2.02&1.52&0.11& 5 			 \\
&RX Boo		& &M7.5III	     &  SRb		  &1.28&3.26&0.18& 5 			 \\
&Egg Nebula& AFGL 2688& F5Iae		     &		  &0.81&13.36&0.22& 6 		\\
&R Hya		& &M7IIIe	     &  Mira		  &1.34&3.18&0.36& 7			 \\
&IRC+10420 &V1302 Aql& F8Ia		     &		  &0.19&2.00&2.96& 3 			 \\
&W Hya		& &M8e		     &  SRa 		  &0.73&34.87&2.49& 7 	 \\
&$\chi$ Cyg	& & S8,K0III	     &	Mira		  &1.29&1.96&4.43& 4		   	 \\
&$\alpha$ Ori	& & M1.5Iab	     &	SRc		  &1.13&6.94&4.84& 5 		 \nl
&$\alpha$ Her	& & M5		     &			  &1.08&7.23&12.11& 5 			 \nl
\\
3a&IK Tau & IRC+10050& M6me		     &  Mira		  &4.79&7.85&12.70& 2 			 \\
&$o$~Cet & Mira    & M7IIIe	     &  Mira		  &3.09&4.99&2.41& 5 \\
&R Aqr		& &M7IIIpe	     &  Mira		  &12.35&3.68&0.60& 5			 \\
&U Her		& &M7III		     &  Mira		  &3.76&14.29&2.27& 4 \\
&U Ori		& &M8III		     &  Mira		  &2.49&0.23&1.11& 4 	 \\
&VX Sgr		& &M4Iae		     &  SRb		  &1.61&2.98&8.79& 5			 \\
&VY CMa		& &M5Iae		     &  SR		  &41.31&7.14&7.39& 5 		 \\
\\
3b&$\alpha$ Sco	& & M1.5Ib	     &  SRa		  &2.83&14.25&10.07& 5			 \nl
&g Her		& &M6III		     &  SRb		  &7.39&2.86&0.82& 5			 \\
&NML Cyg& IRC+40448& {MI\tablenotemark{b}}  &  SR		  &4.80&28.36&3.29&  2\\
&R Leo		& &M8IIIe	     &  Mira		  &10.11&9.29&0.28& 5 			 \\
\\
3c&CIT 6 & RW LMi & C		   &  Mira	  &3.19&15.11&11.81& 2			 \\
&      & IRC+30219 &&&&&&\\
&IRC+10216 & CW Leo &  C & Mira		                  &1.71&1.55&1.99& 2		 \\
&V Hya 		& &C9		     &  SRa		  &3.53&0.76&5.31& 5  	 \\
\enddata

\tablenotetext{a}{From Simbad Database}
\tablenotetext{b}{Supergiant classification based on Morris \& Jura (1983)}

\tablerefs{
1. Gillett, Low \& Stein 1968; 2. Merrill \& Stein 1976b;
3. Giguere, Woolf \& Webber 1976; 4. Forrest, Gillett \& Stein 1975;
5. Merrill \& Stein 1976a; 6. Merrill \& Stein 1976c;
7. Hackwell 1972
}
\end{deluxetable}


\end{document}